\documentclass[fleqn,usenatbib]{mnras}

\usepackage{newtxtext,newtxmath}

\usepackage[T1]{fontenc}

\DeclareRobustCommand{\VAN}[3]{#2}
\let\VANthebibliography\thebibliography
\def\thebibliography{\DeclareRobustCommand{\VAN}[3]{##3}\VANthebibliography}

\usepackage{graphicx}	
\usepackage{amsmath}	
\usepackage{float}

\newcommand{\rubin}{\textit{Rubin}}

\title[ASKAP Monitoring of the Rubin DDFs]{ASKAP Monitoring of the Rubin Deep Drilling Fields}

\author[D. Dobie et al.]{Dougal Dobie$^{1,2}$\thanks{d.dobie@sydney.edu.au},
James Freeburn$^{3,2}$,
Iris de Ruiter$^{1,2}$,
Katie Auchettl$^{4,2}$,
James Coulson$^{5,6}$,
\newauthor
Brent Miszalski$^{5,6}$,
Anais Möller$^{7,2}$,
Nikhil Sarin$^{8,9}$
\\
$^{1}$Sydney Institute for Astronomy, School of Physics, University of Sydney, Sydney, NSW 2006, Australia\\
$^{2}$ARC Centre of Excellence for Gravitational Wave Discovery (OzGrav), Hawthorn, Victoria, 3122, Australia\\
$^{3}$Department of Physics and Astronomy, University of North Carolina at Chapel Hill, Chapel Hill, NC 27599-3255, USA\\
$^{4}$School of Physics, The University of Melbourne, Parkville, VIC 3052, Australia\\
$^{5}$Australian Astronomical Optics, Faculty of Science and Engineering, Macquarie University, NSW, 2109, Australia\\
$^{6}$Astrophysics and Space Technologies Research Centre, Macquarie University, NSW, 2109, Australia\\
$^{7}$ Centre for Astrophysics and Supercomputing, Swinburne University of Technology, John St, Hawthorn, VIC 3122, Australia\\
$^{8}$Kavli Institute for Cosmology, University of Cambridge, Madingley Road, Cambridge, CB3 0HA, UK\\
$^{9}$Institute of Astronomy, University of Cambridge, Madingley Road, Cambridge, CB3 0HA, UK
}

\date{Accepted XXX. Received YYY; in original form ZZZ}

\pubyear{\the\year{}}

\begin{document}
\label{firstpage}
\pagerange{\pageref{firstpage}--\pageref{lastpage}}
\maketitle

\begin{abstract}
In the past decade, time domain astronomy surveys across the multi-messenger spectrum have developed ``alert streams'' which circulate observational information in real-time to the wider astronomical community. Despite this being standard practice for optical, gamma-ray and gravitational wave time domain astronomy for several years now, this kind of infrastructure has not yet been deployed at-scale for radio telescopes. We are conducting a long-term monitoring program that targets the {\it Rubin} Deep Drilling Fields using the Australian SKA Pathfinder (ASKAP). The primary goal of this program is to provide serendipitous radio observations of all objects of interest in these fields, and circulate that information in as close to real-time as possible. In this manuscript we outline the survey plans and status, the infrastructure we have developed to disseminate this information and an initial data release comprising 5 months of ASKAP observations of each field (although only 3 months overlaps with \rubin) conducted on an approximately 10 day cadence with a noise level of $\sim 60\,\mu$Jy. As part of this data release, we present updated high-cadence radio light curves to the tidal disruption event AT2018hyz and the supernova SN2024abfo along with the discovery of two new radio variables, a possible tidal disruption event ASKAP J095711.5+051205 and an in-progress extreme scattering event ASKAP J005026.4-422958. The scientific results from this program along with the infrastructure developed to acquire them will lay the foundation for the integration of future radio imaging surveys into the broader multi-messenger alert stream, leading the way for major facilities such as the Schmidt Observatory System and the SKA.
\end{abstract}


\section{Introduction}
The {\it Vera C. Rubin Observatory} is undertaking the most comprehensive census of the time-variable optical-sky ever conducted -- the Legacy Survey of Space and Time \citep[LSST;][]{2019ApJ...873..111I}. The bulk of {\it Rubin} time will be spent uniformly observing the Southern sky, but five pointings have been selected as ``Deep Drilling Fields'' (DDFs). These fields (COSMOS, ELAIS-S1, XMM-LSS, ECDFS and EDF-S) have been selected based on the large amount of existing archival coverage and the absence of bright optical sources within them. They will be visited $\sim 30$ times\footnote{According to the v5.3.0 baseline observing strategy \citep{lsst_baseline_v53}.} more than any of the standard ``Wide Fast Deep'' (WFD) fields, with the COSMOS field visited twice as often as the other DDFs. The DDF cadence will provide multiple observations every 2 to 3 days in different wavelength bandpasses, surpassing the WFD cadence of every few days and provides a unique opportunity to study ``fast'' transients that evolve on hour-day timescales. These transients, such as kilonovae, on-axis synchrotron afterglows, and Fast Blue Optical Transients will be difficult, but not impossible, to detect in the WFD survey \citep[e.g.][]{2026ApJ...998....8S} but are among the most interesting classes accessible with Rubin.

Rubin will detect up to 10 million variable and transient events per night with a signal-to-noise ratio larger than 5 \citep{ldm612}. However, the sheer volume of sources that {\it Rubin} will discover makes it impossible to conduct comprehensive targeted follow-up of all of them. This problem is particularly obvious at radio wavelengths because only a handful of radio interferometers exist and access to them for transient follow-up is in high demand. Radio follow-up provides unique insight into explosion energetics, the presence of a jet and its properties, and the density of the surrounding medium. This presents a substantial problem in the {\it Rubin} era and necessitates stringent prioritisation of resource usage.

An alternative to the traditional follow-up approach is to conduct complementary widefield surveys that provide coverage of {\it Rubin} discoveries without the need for dedicated observations. This philosophy has been adopted in the past for much smaller surveys \citep[e.g.][]{2020MNRAS.491.5852A,2022MNRAS.516..540R,2023MNRAS.519.4684D,2025MNRAS.538.1800H,2026MNRAS.551g1420M} using quasi-simultaneous coordinated observations. However, this approach monopolises each telescope involved, meaning that it does not scale well to large long-term surveys such as the LSST. The {\it Rubin} scheduling preference for low-airmass observations\footnote{the v5.3.0 baseline observing strategy contains a median airmass of 1.2} substantially limits potential co-observability, while the dynamic survey scheduling presents a fundamental difficulty to any coordination with other facilities. While direct coordination and simultaneous observations may not be feasible, it is still possible to observe the same fields as Rubin at roughly the same times and several optical surveys plan to do so \citep[e.g.][]{2026arXiv260623784R}

In this work we describe a survey on the Australian SKA Pathfinder (project code AS331; PI: Dobie) that targets the DDFs using contemporaneous long-term monitoring that is executed alongside, but independently of, the {\it Rubin} observations. This contemporaneous approach enables the same science goals as coordinated observations, except where true simultaneity is required, while being no more disruptive to telescope operation than any other observing program. Our survey will provide efficient radio coverage of a large number of transient and variable sources discovered by {\it Rubin}, providing high cadence ($\sim 10$\,d) light curves that are ready for scientific interpretation, but also allowing for follow-up resource prioritisation by providing an immediate indicator of which sources have radio counterparts and therefore may warrant X-ray, broadband radio or optical follow-up. 

This is the first program to search for joint optical-radio variability at these scales, and presents an opportunity to probe new parameter space and develop infrastructure for future multi-wavelength coordination. More broadly, these ASKAP observations will form a legacy product that is useful to the wider astronomical community by providing some of the deepest radio observations of these fields ever obtained. In addition to the preliminary data release accompanying this manuscript, we have also built the infrastructure to make the observational and value-added data publicly available via Data Central\footnote{Value-added data will be available from \url{https://apps.datacentral.org.au/askap-rubin-ddf/}.} \citep{Mannering2019} as soon as they pass scientific validation. This service is under active development and adapts the robust automated data transfer and archive ingestion approach described by \citet{Coulson2026}. We will continue to improve this infrastructure in response to community demand.

\begin{figure}
    \centering
    \includegraphics[width=\linewidth, trim={0.3cm 0.5cm 0cm 1cm}, clip]{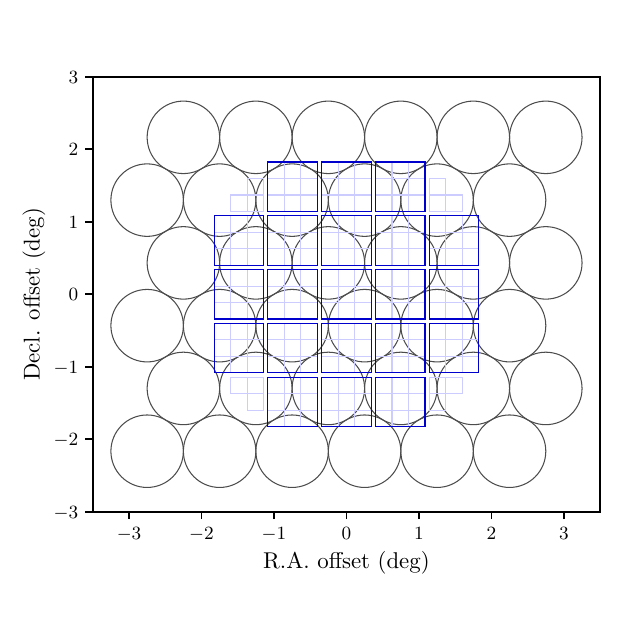}
    \caption{Comparison of the {\it Rubin} (blue; showing the footprint of each CCD and ``raft'') and ASKAP (grey; showing beam FWHMs) footprints for single pointings centered on the same location. The ASKAP images extend beyond the beam FWHM, with each beam mosaicked together, while the {\it Rubin} pointings are dithered to fill in the chip-gaps.}
    \label{fig:relative_footprints}
\end{figure}

\begin{figure*}
    \centering
    \includegraphics[]{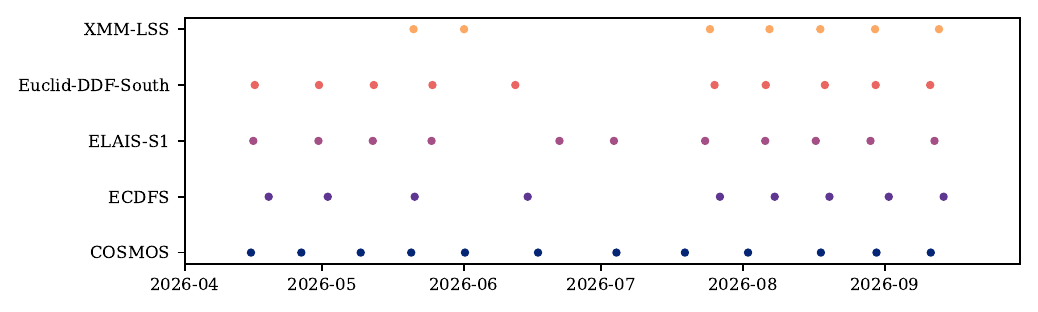}
    \caption{Dates for all observations executed as part of this survey. While there were some issues with the execution of early observations, and some disruption to observing from mid-June, we have achieved an observing cadence of $\sim 12$ days from mid-July onwards. This is close to the nominal survey cadence of 10 days, despite the scheduler executing these observations as ``filler''.}
    \label{fig:survey_progress}
\end{figure*}

\section{Survey and Data}
\subsection{Observing Strategy}
The Australian Square Kilometre Array Pathfinder \citep[ASKAP;][]{2021PASA...38....9H} is a radio interferometer consisting of thirty-six 12\,m antennas located at Inyarrimanha Ilgari Bundara, the CSIRO Murchison Radio-astronomy Observatory, in Western Australia. Each antenna is equipped with a Phased Array Feed (PAF) that can operate between 0.7--1.8\,GHz, giving a total of 36 beams. In this work we use the same PAF configuration as the Evolutionary Map of the Universe \citep[EMU;][]{2025PASA...42...71H} survey, consisting of the {\sc closepack36} PAF footprint with 288\,MHz of bandwidth centered on 943\,MHz. This ultimately provides an approximately square field of view, sufficient to comfortably enclose the entire footprint of a single \rubin\ pointing as shown in Figure \ref{fig:relative_footprints}. 

This program targets the five \rubin\ Deep Drilling Fields \citep[DDFs;][]{2024ApJS..275...21G}, with ASKAP pointings centered on the respective DDFs, except for the Euclid Deep Field South field, which is covered by two Rubin pointings and by a single ASKAP pointing located at 04:04:57.84 $-$48:25:22.8 that covers the bulk of the footprint. We selected the Deep Drilling Fields for two main reasons. First, they are all legacy fields with comprehensive multi-wavelength coverage. Second, they are observed at a higher cadence than other fields making the prospects for finding rapidly evolving relativistic optical transients much stronger. Additionally, these are the first fields observed by \rubin\ and will also be observed at an even higher cadence during the initial stage of \rubin\ operations, making this program exceedingly timely. We analyse the data for the entire ASKAP footprint rather than restricting ourselves to the \rubin\ footprint.

The canonical evolution timescale for synchrotron afterglows at radio wavelengths is weeks--months. Hence, the optimal observing strategy for targeting these sources would be deep observations every few weeks. However, these observations make use of ``filler'' time (gaps in the observing schedule where the telescope would otherwise be sitting idle) similar to the Commensal Real-time ASKAP Fast Transient survey \citep{2025PASA...42...36S}. Unlike other projects \citep[e.g.][]{2026arXiv260623784R}, our observations are executed independent of \rubin\ observing or any other multi-wavelength triggers. We adopt a pragmatic observing strategy of 3\,h observations, achieving a sensitivity of $\sim 60\,\mu$Jy on a cadence of approximately 10 days. 
This makes it easier to fit our observations into the schedule and also provides the ability to recover EMU-like sensitivity on a monthly cadence by stacking observations. This increased cadence also comes with the positive side effect of being able to rapidly determine source evolution and easily distinguish between real synchrotron afterglows and contaminants such as stochastic variability from Active Galactic Nuclei (AGN), along with increased sensitivity to shorter variability timescales.

This program began observing on 2026-04-15 and in this paper we report the 49 observations obtained through to 2026-09-13. This represents 30\% of the approved survey, which consists of 33 observations of each field through to April 2027. As an auxiliary science output, we plan to combine all the data we obtain into a single deep image of each field which should achieve a $\sim 10\,\mu$Jy sensitivity, comparable to the planned EMU sensitivity \citep[][]{2011PASA...28..215N}, 2--3 times better than the actual EMU sensitivity \citep[][]{2025PASA...42...71H} and among the deepest radio images of these fields ever obtained. Figure \ref{fig:survey_progress} illustrates how the survey has been executed so far, showing that we achieve close to the nominal 10 day cadence, other than during periods of general observing disruption. Based on the success of the survey so far, we plan to extend the survey beyond the initial request and hope to provide ongoing coverage of these fields across the lifetime of the LSST.

\subsection{Additional Observations}
ASKAP began science observations by conducting several pilot surveys from 2019 through to 2022. Since beginning full operations in 2022, ASKAP has been predominantly dedicated to conducting six Survey Science Projects (SSPs), with the remainder of the time spent conducting observatory-led projects (such as the Rapid ASKAP Continuum Survey) and, as of 2024, Guest Science Projects. All ASKAP data is made publicly available after it passes scientific validation, and hence there is a moderate amount of serendipitous coverage of the DDFs across the past seven years. In this work we also include any publicly available ASKAP imaging that overlaps with the footprint of the monitoring program fields. We will also continue to monitor the archive for new serendipitous observations and add them to our dataset as they become available. Of note, the Variables And Slow Transients (VAST) survey science project covers four of the deep drilling fields as part of its Extragalactic survey \citep{2026PASA...43...39D}, meaning that these fields are covered by shallow observations on a two-month cadence. The dynamic radio sky is not well-characterised, so having reference imaging with a long time baseline makes real-time classification of candidate radio transients substantially easier. 

In addition to the serendipitous archival data, we also obtained deep ASKAP imaging of the COSMOS ($2\times 9$\,h), ELAIS-S1 (10\,h) and ECDFS (10\,h) between June and September 2024 under an ASKAP Guest Science Project (AS310, PI: Dobie).

\subsection{Data processing, validation and quality assurance}
All data obtained as part of this survey are reduced using the {\sc ASKAPsoft} processing pipeline \citep[][]{2019ascl.soft12003G} using the standard continuum imaging parameters comparable to those used to process EMU data \citep{2025PASA...42...71H}. This includes running sourcefinding and generating auxiliary maps using {\sc Selavy} \citep{2012PASA...29..371W}. The data obtained serendipitously come from a variety of surveys conducted with a range of observing and processing parameters and the data quality may vary correspondingly. While all data has passed scientific validation by the relevant survey team, these validation criteria may vary depending on the survey needs. We also note that each observation is validated as a whole, meaning that in some cases, large regions of images may not be of sufficient quality to use for science. We plan to reprocess all relevant observations with uniform processing parameters using the improved {\sc Flint} pipeline\footnote{\url{https://github.com/flint-crew/flint}} (Galvin et al. in prep.), which will comprise a future second data release for this survey.

A key issue with the default ASKAP data products is the incorrect decomposition of bright compact sources into ``islands'' consisting of multiple Gaussian ``components''. This has two key impacts -- the measured properties of any of these components are nonsensical when taken individually and, because the decomposition varies depending on the image properties, the association of these components across observations is not consistent, which results in apparent high-amplitude variability. In the past, these sources (typically bright AGN) have simply been excluded from variability searches through source filtering. Noting the potential application of this dataset for studies of multiwavelength AGN variability, we have attempted to mitigate this by searching for these source finding artefacts using two techniques. \citet{2024PASA...41....3D} account for this by searching for multiple-component islands and satisfying
\begin{equation}
    \frac{ab}{\theta_M\theta_m} < {\rm median}\left(\frac{ab}{\theta_M\theta_m}\right)
\end{equation}
where $a$ and $b$ are the major and minor axes of the source (or in this instance, the island defined by {\sc Selavy}) respectively, $\theta_M$ and $\theta_m$ are the major and minor axes of the image PSF and the right term is the median of that quantity across the source sample. This approach is effective in removing sources that are split into multiple components due to noise spikes in the vicinity of the source. We also apply a second filter, motivated by truly variable sources being necessarily compact on these angular scales, searching for sources divided into multiple components where the maximum distance between any two components is less than one beam-width. Some truly extended sources will likely pass this criteria, but will be removed from any transient searches anyway. For each image used in this project we update the flux density, position and size measurements of all objects satisfying either filter by fitting a single Gaussian to collapse them into a single component.

Additionally, for all archival observations we crop the {\sc Selavy} catalogues to a $7.3\times 6.3\,\deg$ region located on the centre of the relevant ASKAP-DDF field such that our dataset only includes objects within the footprint of our survey's observations.

\subsection{Source association and statistics}
All ASKAP dataproducts become available on CASDA immediately after scientific validation, which generally involves assessing image quality metrics alongside manual inspection (details vary between projects). We have developed a listener script to monitor CASDA for new observations covering the regions of interest and immediately download and apply the processing described above. This script also downloads and processes all AS331 observations prior to scientific validation in order to minimise latency. We then run the data through the VAST pipeline\footnote{\url{https://vast-survey.org/vast-pipeline/}} \citep[][]{vast_pipeline} which carries out source association, measures forced photometry of non-detections and calculates variability metrics. This pipeline has been used extensively in the literature and we refer the reader to \citet{2022MNRAS.510.3794D} for details of its use.

In this work we use a novel approach to running the VAST pipeline, employing two separate pipeline runs, one using the data obtained as part of this survey, and a second that includes all relevant data on CASDA. Previous work \citep[e.g.][]{2025MNRAS.538.2676G,2026PASA...43...39D} uses a single pipeline run that includes all the survey data, and then separately queries the relevant archives for other observations. Our new approach is faster overall (obtaining all data for a single source in CASDA can take minutes) and also more robust as we are able to calculate variability metrics for all sources in our sample. However, this approach is only possible due to the relatively small sky coverage involved -- scaling to truly widefield surveys would require robust, dedicated infrastructure in order to meet the storage and computational power requirements. This infrastructure would be feasible to build as a community resource (comparable to the Rubin brokers), but not for individual researchers or small groups. Looking ahead to the era of the SKA Observatory, we strongly recommend that this functionality is built-in to the image-domain transient pipelines that are deployed as part of the SKA Regional Centre Network.

\subsection{Data Format and Access}
The VAST pipeline produces a series of parquet files that contain all database tables for each pipeline run. This data can then be accessed programmatically using {\sc vast-tools}\footnote{\url{https://vast-survey.org/vast-tools/}} \citep{vast_tools}, which provides a straightforward way to search for objects of interest using pandas queries (along with the capacity for more complex filtering), produce cutouts and light curves. To make this data generalisable we strip the paths for all dataproducts, leaving only the filename, such that users can organise the dataset on their own machines as they see fit.

We will provide a live data release via Data Central \footnote{\url{https://apps.datacentral.org.au/askap-rubin-ddf/}} with access to these tables, which is updated as new observations become available -- this service is under active development and more features (e.g. direct access to the ASKAP dataproducts, a cutout service and a Kafka stream) will become available as the survey progresses. The VAST pipeline output used in this paper is available at \url{https://doi.org/10.5281/zenodo.23224747}, along with an example {\sc Jupyter} notebook that demonstrates how to interact with the data. This notebook also shows other capabilities such as linking pipeline runs together and calculating per-band variability metrics that are not covered by the standard {\sc vast-tools} documentation.

\begin{table}
\begin{tabular}{lllr}
\hline\hline
   Field &  SBID &    Start Time (UTC) & $\sigma_{\rm med}$ ($\mu$Jy) \\
\hline
\hline
  COSMOS & 83773 & 2026-04-15 10:51:36 &                           62 \\
         & 83946 & 2026-04-26 10:59:52 &                           72 \\
         & 84319 & 2026-05-09 10:35:32 &                           75 \\
         & 84611 & 2026-05-20 11:25:21 &                           73 \\
         & 85242 & 2026-06-01 05:26:26 &                           66 \\
         & 85673 & 2026-06-17 04:37:52 &                           60 \\
         & 86293 & 2026-07-04 07:54:14 &                           55 \\
         & 86766 & 2026-07-19 06:57:48 &                           65 \\
         & 87177 & 2026-08-02 02:15:15 &                           61 \\
         & 87601 & 2026-08-18 00:16:43 &                           78 \\
         & 87880 & 2026-08-30 03:59:04 &                           96 \\
         & 88233 & 2026-09-11 00:44:05 &                           76 \\
\hline
   ECDFS & 83866 & 2026-04-19 07:48:03 &                           59 \\
         & 84127 & 2026-05-02 05:45:26 &                           56 \\
         & 84627 & 2026-05-21 05:39:22 &                           60 \\
         & 85629 & 2026-06-14 22:22:56 &                           51 \\
         & 86990 & 2026-07-26 22:45:14 &                           51 \\
         & 87344 & 2026-08-07 22:05:14 &                           47 \\
         & 87643 & 2026-08-19 21:06:53 &                           48 \\
         & 87949 & 2026-09-01 20:24:47 &                           51 \\
         & 88302 & 2026-09-13 20:00:55 &                           50 \\
\hline
ELAIS-S1 & 83783 & 2026-04-15 23:10:29 &                           64 \\
         & 84086 & 2026-04-30 04:43:14 &                           68 \\
         & 84396 & 2026-05-12 01:51:13 &                           60 \\
         & 84720 & 2026-05-24 22:04:02 &                           57 \\
         & 85703 & 2026-06-21 20:53:03 &                           52 \\
         & 86291 & 2026-07-03 18:46:49 &                           49 \\
         & 86886 & 2026-07-23 17:09:05 &                           56 \\
         & 87272 & 2026-08-05 20:09:55 &                           59 \\
         & 87565 & 2026-08-16 21:27:41 &                           68 \\
         & 87858 & 2026-08-28 20:32:10 &                           60 \\
         & 88273 & 2026-09-11 20:15:00 &                           61 \\
\hline
    EDFS & 83803 & 2026-04-16 06:52:28 &                           55 \\
         & 84087 & 2026-04-30 07:44:33 &                           55 \\
         & 84407 & 2026-05-12 07:13:29 &                           52 \\
         & 84729 & 2026-05-25 02:50:44 &                           50 \\
         & 85610 & 2026-06-12 05:29:06 &                           52 \\
         & 86946 & 2026-07-25 18:36:06 &                           51 \\
         & 87273 & 2026-08-05 23:13:23 &                           49 \\
         & 87626 & 2026-08-18 21:05:56 &                           49 \\
         & 87877 & 2026-08-29 23:37:48 &                           55 \\
         & 88232 & 2026-09-10 21:39:51 &                           51 \\
\hline
 XMM-LSS & 84621 & 2026-05-20 23:47:18 &                           62 \\
         & 85235 & 2026-06-01 00:45:26 &                           63 \\
         & 86906 & 2026-07-24 18:26:49 &                           74 \\
         & 87303 & 2026-08-06 18:38:27 &                           61 \\
         & 87600 & 2026-08-17 21:13:34 &                           67 \\
         & 87876 & 2026-08-29 20:33:40 &                           69 \\
         & 88287 & 2026-09-12 19:48:07 &                           60 \\
\hline\hline
\end{tabular}
\caption{AS331 observations included in this work, showing the start time and the median noise across the image.}
\end{table}

\begin{figure*}
    \centering
    \includegraphics{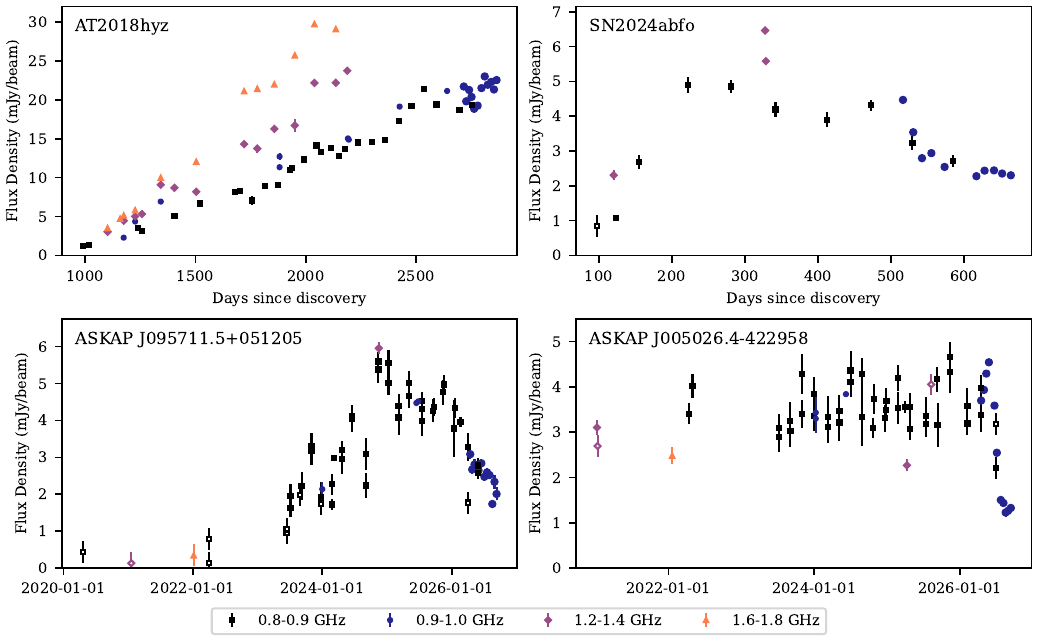}
    \caption{Radio light curves of sources detected in the searches described in Section \ref{sec:searches}. Closed markers denote catalogued sourcefinder measurements while open markers denote forced fits. The light curves of AT2018hyz and SN2024abfo have been truncated to remove prior non-detections. The light curve of AT2018hyz contains dedicated monitoring of the source using the VLA, presented in \citet{2022ApJ...938...28C} and \citet{2026ApJ...998..111C}. Observations from this survey are denoted by larger markers.}
    \label{fig:lcs}
\end{figure*}

\section{Preliminary Searches}
\label{sec:searches}
We present three searches for objects of particular interest of the survey data obtained to date. This dataset comprises 49 observations obtained under AS331 along with 879 archival observations with overlap of the DDFs, most of which come from the VAST survey. Details of these observations are available in the supplementary material. We present two searches that are led by multi-wavelength information, crossmatching to the Transient Name Server and Rubin data, while the third is a radio-driven search. These searches are not designed to be exhaustive, but rather demonstrate the utility of this dataset. 

\subsection{Transient Name Server crossmatch}
We searched for radio counterparts to known multi-wavelength transients within the footprint of our ASKAP survey by crossmatching our source catalogue to the TNS source catalogue with a radius of 5''. This resulted in 214 candidates with a maximum signal-to-noise ratio greater than 8, but without applying any quality cuts. We then manually inspected all candidates and after removing artefacts and non-variable radio sources (likely either the optical transient being a variable AGN or occurring in a radio-bright host galaxy), we were left with two optical transients with viable radio counterparts. Neither of these are within the Rubin DDFs, but are covered by the footprint of our ASKAP DDF monitoring program (see Fig. \ref{fig:relative_footprints}).

\subsubsection{AT2018hyz}
AT2018hyz is a Tidal Disruption Event (TDE) discovered by the All-Sky Automated Survey for Supernova (ASAS-SN) at a redshift of $z=0.04573$ \citep{2020ApJ...903...31H,2020MNRAS.498.4119S,2020MNRAS.497.1925G}. This TDE is notable as it is among an emerging class of events exhibiting late-time radio flaring, with radio emission not detected until three years post-discovery. Due to this novelty, it is well-studied at radio wavelengths \citep[][]{2024ApJ...974..241A,2026ApJ...998..111C}. Figure \ref{fig:lcs} shows the radio light curve of AT2018hyz created from publicly available ASKAP data, along with the measurements reported by \citet{2022ApJ...938...28C,2026ApJ...998..111C} at frequencies below 1.8\,GHz. We find broad agreement between the ASKAP light curve and the contemporaneous literature light curve, while our more recent data suggests the radio emission has flattened, and potentially peaked, at $\sim 2750$ days post-discovery. The AS331 data exhibits a large amount of scatter and using the formalism of \citet{2023MNRAS.521..389R} we measure a fractional variability of $6.6 \pm 0.1$\,\%. This scatter is not obvious in the earlier light curve due to the lower observing cadence. We tentatively rule out telescope systematics as the origin of this scatter as no other objects in the field show comparable variability, leaving two potential explanations. The observed variability may be intrinsic \citep[such as the scenario proposed for the TDE AT2022cmc by][]{2023MNRAS.521..389R} or arising from scintillation due to plasma along the line of sight.

If it is induced by scintillation, the source must remain reasonably compact at late times. Using the approach proposed in \citet{2020MNRAS.494.2449D}, the expected level of refractive scintillation along this line of sight is $\sim 27\%$ with a timescale of 10\,days. Using the equations of \citet{1998MNRAS.294..307W}, the lower-than-expected variability implies that the source is larger than the refractive scintillation scattering disk by a factor of $(F_{var,{\rm exp}}/F_{var,{\rm obs}})^{6/7} = 3.4$. The variability timescale also increases by a factor of $F_{var,{\rm exp}}/F_{var,{\rm obs}}$ to $\sim 40$\,days, consistent with our observations. We therefore infer an approximate source size of $\sim 2$\,mas, or a physical size of $\sim 5\times 10^{18}$\,cm which is in broad agreement with the delayed outflow modelling from \citet{2026ApJ..1007..103M} and high resolution radio imaging from Golay et al. (in prep.).

\begin{figure*}
    \centering
    \includegraphics{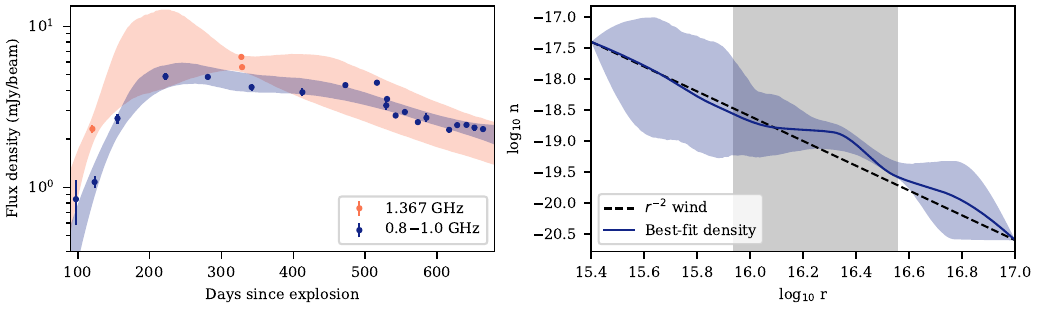}
    \caption{Modelling of the radio emission from SN2024abfo, as described in Section \ref{subsubsec:sn2024abfo}. Left: Light curve with the 90\% interval of the model fit shaded. Right: The inferred density profile. We show the best fit model with 90\% interval shaded along with a reference $r^{-2}$ wind. We also shade the part of the parameter space that the ASKAP observations are sensitive to in grey.}
    \label{fig:sn2024abfo_model}
\end{figure*}

\subsubsection{SN2024abfo}
\label{subsubsec:sn2024abfo}
SN2024abfo is a type IIb supernova discovered by the Asteroid Terrestrial-impact Last Alert System in the nearby galaxy NGC\,1493 \citep{2025A&A...698A.129R,2025A&A...704A..89D}. The radio counterpart to SN2024abfo was first reported by \citet{2024ATel16920....1R} with a 9\,GHz flux density of $0.36\pm 0.03$\,mJy at 6 days post-explosion. \citet{2026PASA...43...39D} reported serendipitous observations of the supernova as part of the VAST extragalactic survey, with a single detection $\sim 4$ months post-explosion.

Figure \ref{fig:lcs} shows the radio light curve of SN2024abfo created from publicly available ASKAP data. 
We fitted the 22 ASKAP detections at 0.8-0.9 and 1.3\,GHz with a model for synchrotron emission from the interaction of supernova ejecta with pre-existing circumstellar material (CSM) via \texttt{redback-csm}~\citep{Sarin2024, Sarin2026}. We model the CSM as a static density profile represented by 6 nodes at fixed radii. The nodes are allowed to move freely in density while ensuring an overall smooth density profile. The radio spectrum includes a smooth synchrotron self-absorption and an additional, time-dependent opacity with a $\nu^{-2.1}$ dependence, as expected for free-free absorption. We allowed both ejecta mass and explosion energy to vary; the ejecta-mass range of $1$--$10\,M_{\odot}$ encompasses the optical estimate of $4.1\pm0.7\,M_{\odot}$ and the
possible additional $1$--$2\,M_{\odot}$ in outer shells discussed by
\citet{2025A&A...698A.129R}. We fixed the electron index to $p=3$ and the energy fractions to $\epsilon_e=0.1$ and $\epsilon_B=0.01$ for simplicity. A 10\% fractional flux-scatter term was added in quadrature to the quoted uncertainties. 
As shown in Figure \ref{fig:sn2024abfo_model}, the spline-density model with additional opacity reproduces both ASKAP frequency tracks, with a density profile (bottom panel) closely following a power-law with minor departures.
Exploratory fits with a simple $r^{-2}$ density (i.e., a constant velocity wind) or without the additional opacity leave larger residuals. The fitted effective free-free optical depth peaks near $\tau_{\rm ff}(1\,\mathrm{GHz})\simeq0.9$ at approximately 350\,d. The radio-only ejecta-mass posterior is broad, with a 90\% interval of $1.4$--$9.1\,M_{\odot}$.

For the maximum-posterior solution, the shock crosses radii of $0.86\times10^{16}$ to $3.62\times10^{16}$\,cm during the ASKAP
observations. If the reconstructed density arose from a spherical, constant-velocity outflow at $v_{\rm wind}=20\,\mathrm{km\,s^{-1}}$, these radii correspond to material ejected approximately 140--570\,yr before explosion, with $\dot M=4\pi r^2\rho v_{\rm w}$ reaching approximately $2.5\times10^{-5}\,M_{\odot}\,\mathrm{yr^{-1}}$ near 350\,yr. 
Integrating over all our posterior samples own shock-radius interval gives a CSM mass of $0.0084\,M_{\odot}$ (90\% interval
$0.0071$--$0.0113\,M_{\odot}$). 
The reconstructed non-wind density profile suggests a varying pre-explosion outflow, consistent with the partially stripped progenitor, although its precise timing
and the mass-loss mechanism are not established by these radio data.
The inferred rate and lookback times depend on the assumed wind speed and radio microphysics.


\subsection{Radio variability search}
\label{subsec:radio-variability}
Additionally, we searched for variable radio sources independent of any multi-wavelength information. We conducted two standard $\eta$--$V$ searches - the first used only the data obtained under AS331. For the second search we used the pipeline run that uses all CASDA data, but recalculated the variability metrics to only include ``deep'' observations with observing frequencies below 1\,GHz. Functionally, this means excluding data from the VAST, RACS and FLASH surveys. For both searches we applied a quality filter comparable to previous work \citep[e.g.][]{2024MNRAS.535..909D,2026PASA...43...39D}, requiring that all sources satisfy
\begin{itemize}
    \item \texttt{max\_snr} > 10 and \texttt{avg\_compactness} < 1.4
    \item \texttt{n\_measurements} > 5 and \texttt{n\_selavy} > 1
    \item \texttt{n\_neighbour\_distance} > 0.5\,'
    \item \texttt{n\_relations} = 0 and \texttt{n\_siblings} = 0
\end{itemize}
We initially used the same $\eta$--$V$ approach as \citet{2022MNRAS.510.3794D}, where we consider the $\eta$ and $V$ distributions independently and define sources with $\eta > \mu_\eta > 1.5\sigma_\eta$ and $V > \mu_V + 1.0\sigma_V$ to be transient candidates, corresponding to cutoffs of $\eta > 21$ and $V>0.2$. This produced a large number of scintillating sources, so we increased our modulation index cutoff to $V>0.3$. We then inspected the complete light curves and image cutouts for all candidates. All sources that passed these selection cuts appear to be real sources of radio emission, but the bulk of them are qualitatively consistent with stochastic variability after visual inspection. In addition to the objects already discussed, we detect two further sources of interest.

\subsubsection{ASKAP\,J095711.5+051205}
Figure \ref{fig:lcs} shows the light curve of ASKAP\,J095711.5+051205, which is not detected by ASKAP until July 2023, when it exhibits a 2-year rise to a peak in late 2024 and a gradual decay from late 2025. The radio source is spatially coincident with the nucleus of the galaxy Z 36-7, at redshift $z=0.02156$ \citep[][]{2015ApJS..219...12A}. The corresponding radio luminosity of $6\times 10^{28}$\,erg\,s$^{-1}$\,Hz$^{-1}$ is broadly consistent with both a thermal TDE and a weak AGN. The optical spectrum available from the Sloan Digital Sky Survey \citep[][]{2015ApJS..219...12A} shows strong Balmer absorption lines, consistent with a poststarburst galaxy. \citet[][]{2026ApJ..1006...73H} classify the source as having a misaligned central contiguous poststarburst region.

We searched the VLA Sky Survey \citep[VLASS;][]{2020PASP..132c5001L} for radio emission at this position. The source is not detected in VLASS epochs 1 and 2, but is detected in epochs 3 and 4 at approximately 1.2\,mJy. Comparing the VLASS and ASKAP data, we estimate a spectral index of $\sim 1.1$ around February 2023 and $\sim -0.7$ around November 2025. These early radio non-detections constrain the flux density to be over an order of magnitude fainter than at the peak, suggesting that the observed radio variability does not arise from stochastic variability or an intrinsic AGN flare. Considering the host galaxy classification, early non-detections, spectral evolution and peak luminosity, we propose that this object is a TDE with no known optical counterpart. Given the uncertain event time, under this interpretation it is not possible to determine whether the lack of optical counterpart arises from intrinsically faint emission or was simply not at observed at the time of the event.

\subsubsection{ASKAP\,J005026.4$-$422958}
Figure \ref{fig:lcs} shows the lightcurve of ASKAP\,J005026.4$-$422958, showing a reasonably stable flux density of $\sim 3.5\,$mJy from early 2021 through to early 2026 followed by a sharp spike and rapid decline. This morphology is qualitatively consistent with an in-progress Extreme Scattering Event. Ongoing monitoring as part of AS331 will test this interpretation, where we should see another sharp spike and then a return to a relatively stable flux density as the source emerges from behind the plasma lens. 

\subsection{Rubin crossmatch}
We searched for sources exhibiting both optical variability detected by Rubin, and radio variability detected by this survey, using two different approaches. We do not consider AGN variability -- both searches reveal too many of them to discuss individually, none appear to be particularly outstanding relative to the others, and a population analysis is beyond the scope of this work. 

The first involved querying \rubin\ public alerts via the Fink broker \citep{2021MNRAS.501.3272M} using their data transfer service\footnote{\url{https://lsst.fink-portal.org/download}}. Fink provides access to public \rubin\ alerts, which are issued for $\geq5\sigma$ detections of sources that have varied in brightness relative to a reference template in difference-image photometry. We queried a list of 31,142 radio sources that pass the quality criteria defined in Section \ref{subsec:radio-variability} and used the inbuilt crossmatching service with a crossmatch radius of $5''$, but did not supply any further filters. We then searched for contemporaneous variability by manually inspecting all sources where the latest Rubin datapoint was observed after the earliest AS331 datapoint. We find multiple sources that exhibit variable optical emission alongside static radio emission, but no sources that are noteworthy variables at both wavelengths.

The second approach was to match to the Vera C. Rubin Observatory Data Preview 2 \citep[DP2;][]{RTN-115,10.71929/rubin/3382528}, which is a release of commissioning data from \textit{Rubin} and is the facility's first data release using LSSTCam \citep{10.71929/rubin/2571927}. It comprises observations spanning $\sim3000$~deg$^2$ over $\sim9$~months between April 2025 and January 2026. The sky coverage comprises areas of the WFD footprint and the DDFs. We query the \texttt{DiaSource} table for DDF sources satisfying the following conditions:

\begin{itemize}
    \item \texttt{snr} > 10
    \item \texttt{isDipole} = False
    \item \texttt{shape\_flag} = False
    \item \texttt{glint\_trail} = False
    \item \texttt{psfFlux\_flag} = False
\end{itemize}

This yielded 266,815 unique DIA (Difference Image Analysis) objects of which 1,452 had a radio source cross-match.  We constructed their forced photometry light curves from the \texttt{ForcedSourceOnDiaObject} table and then manually inspected each light curve. This approach also produced no sources that are particularly noteworthy, with low variability amplitude, no clear trend in the variability or broadly consistent with AGN variability.

While these approaches have not yielded any sources of interest, they are limited by two factors -- the AS331 survey currently contains less than five months of observations, (relatively short compared to the timescale of synchrotron afterglow evolution) and at the time of submission, Rubin has been offline since 15 July 2026 due to storm damage so there are only two months of contemporaneous observing. Future iterations of this approach will be more productive as both surveys dramatically expand.

\section{Discussion and Conclusions}
As part of this survey we have detected two new highly variable radio sources and provided additional coverage of two known radio afterglows. The discoveries and insights gained from this data were only possible due to the wide field of view (46\,sq.\,deg. per field), good sensitivity ($\sim 60\,\mu$Jy) and relatively high cadence ($\sim 2$ weeks), the combination of which sets this work apart from all previous radio time domain imaging surveys. The cadence is particularly noteworthy, as it allows us to measure the scintillation properties of AT2018hyz and infer the physical size of its outflow, and also characterise the rapid evolution of ASKAP J005026.4-422958. Despite the utility of these kinds of measurements, no existing survey has observed on these timescales and even future surveys on facilities such as the Deep Synoptic Array are being planned with months-long cadences \citep[][]{2026ApJ..1007..192F}.

The results presented in this manuscript are the tip of the iceberg, both in the sense of the size of the analysed dataset relative to the scale of the planned surveys, and in terms of the potential utility of these surveys in the hands of the wider astronomical community. Our work is limited to 3 months of contemporaneous observing which gives a limited temporal baseline with which to analyse multi-wavelength variability, but also means that the radio afterglows of many optical transients will not yet have peaked, and may not have even emerged \citep[e.g.][]{2024ApJ...971..185C}. The search and analysis presented in this work are intentionally lightweight, prioritising rapid turnaround, and this dataset presents a ripe opportunity for the wider astronomical community to pursue a variety of science cases in greater depth.

In addition to the direct results presented in this work, this program represents two substantial advancements in time domain astronomy. 

This is the first time that a radio transient imaging survey has produced publicly available results in near-real-time. While several FRB search pipelines produce near-real-time alerts, the computational overhead associated with forming radio images has meant that imaging searches (sensitive to variability on $\gtrsim$ minute timescales) have had large amounts of human input and the delay between observations and public results has been $\sim 2$ months at best \citep[][]{2019ApJ...887L..13D}, but may be as slow as years. The process of scheduling an observation through to producing publicly available images has been automated on ASKAP for some time, and as part of this work we build upon that by monitoring the ASKAP archive for new observations and automatically downloading, processing and aggregating the data to produce light curves. There is substantial latency in the overall dataflow (the largest being the initial data reduction), but it is now fully automated with no human-in-the-loop and updated light curves are made accessible as soon as new data becomes available. In practice, the data is processed and made available on the ASKAP archive within a few days of observation and light curves are available within tens of minutes from there. 

In the broader context of multi-messenger astronomy, this is the first time that it has been possible to link data streams from major optical and radio facilities together quasi-automatically. This program operates as a proof of concept for integrating radio data into the global multi-messenger alert stream. While we do not, at the time of submission, release {\sc VOEvent} alerts comparable to optical, gamma-ray or gravitational wave facilities, this capability is under development and will eventually be deployed as part of this survey. Next-generation radio facilities such as the SKAO and the Deep Synoptic Array have plans to release alerts but the infrastructure for doing so does not yet exist. Our real-time data stream presents an opportunity for these facilities to assess community demand and specific infrastructure requirements. Similarly, future facilities also plan to combine multi-messenger data streams \citep[e.g. the Schmidt Observatory System will link the Argus Array and DSA;][]{2026ApJ..1007..192F} to form a holistic picture of the dynamic sky and this work is the first key step towards doing so.

We strongly encourage community feedback on the data products and infrastructure presented as part of this paper, along with input on potential future upgrades.

\section*{Data Availability}
All radio data underlying this project are available from the CSIRO ASKAP Science Data Archive, CASDA (\url{http://data.csiro.au}). The VAST pipeline output for the dataset used in Section \ref{sec:searches}, along with a Jupyter notebook detailing how to use it, is available from \url{https://doi.org/10.5281/zenodo.23224747}. Please see \url{https://apps.datacentral.org.au/askap-rubin-ddf/} for the latest information about the availability of live data products.

\section*{Acknowledgements}
We thank Adam Deller and Elaine Sadler for useful discussions. 

This research was supported by the Australian Research Council Centre of Excellence for Gravitational Wave Discovery (OzGrav), project number CE230100016. This research was supported by the Sydney Informatics Hub, a Core Research Facility of the University of Sydney. This paper makes use of services and code that have been provided by AAO Data Central  (datacentral.org.au). This work was supported by software and data engineering support resources awarded under the Astronomy Data and Computing Services (ADACS) Merit Allocation Program. ADACS is funded from the Astronomy National Collaborative Research Infrastructure Strategy (NCRIS) allocation provided by the Australian Government and managed by Astronomy Australia Limited (AAL). AM is supported by the ARC Discovery Early Research Award (DE230100055).

This scientific work uses data obtained from Inyarrimanha Ilgari Bundara, the CSIRO Murchison Radio-astronomy Observatory. We acknowledge the Wajarri Yamaji People as the Traditional Owners and native title holders of the Observatory site. CSIRO’s ASKAP radio telescope is part of the Australia Telescope National Facility (https://ror.org/05qajvd42). Operation of ASKAP is funded by the Australian Government with support from the National Collaborative Research Infrastructure Strategy. ASKAP uses the resources of the Pawsey Supercomputing Research Centre. Establishment of ASKAP, Inyarrimanha Ilgari Bundara, the CSIRO Murchison Radio-astronomy Observatory and the Pawsey Supercomputing Research Centre are initiatives of the Australian Government, with support from the Government of Western Australia and the Science and Industry Endowment Fund. This paper includes archived data obtained through the CSIRO ASKAP Science Data Archive, CASDA (http://data.csiro.au). This research has made use of the CIRADA cutout service at URL cutouts.cirada.ca, operated by the Canadian Initiative for Radio Astronomy Data Analysis (CIRADA). CIRADA is funded by a grant from the Canada Foundation for Innovation 2017 Innovation Fund (Project 35999), as well as by the Provinces of Ontario, British Columbia, Alberta, Manitoba and Quebec, in collaboration with the National Research Council of Canada, the US National Radio Astronomy Observatory and Australia’s Commonwealth Scientific and Industrial Research Organisation.

This material is based upon work supported in part by the National Science Foundation through Cooperative Agreements AST-1258333 and AST-2241526 and Cooperative Support Agreements AST-1202910 and 2211468 managed by the Association of Universities for Research in Astronomy (AURA), and the Department of Energy under Contract No. DE-AC02-76SF00515 with the SLAC National Accelerator Laboratory managed by Stanford University. Additional Rubin Observatory funding comes from private donations, grants to universities, and in-kind support from LSST-DA Institutional Members. This publication is based in part on proprietary Rubin Observatory Legacy Survey of Space and Time (LSST) data, and was prepared in accordance with the Rubin Observatory data rights and access policies. All authors of this publication meet the requirements for co-authorship of proprietary LSST data. This research uses services or data provided by the Rubin Science Platform at NSF-DOE Vera C. Rubin Observatory, which is jointly funded by the U.S. National Science Foundation and the U.S. Department of Energy, Office of Science. This work was developed within the Fink community and made use of the Fink resources. Fink is supported by LSST-France and CNRS/IN2P3.

\bibliographystyle{mnras}
\bibliography{bibliography}

\bsp	
\label{lastpage}
\end{document}